# Nanoscale mechanics and ultralow Friction of natural 2D silicates: Biotite and Rhodonite

*Surbhi Slathia[a], Manoj Tripathi[b,c*], Raphael Benjamim de Oliveira[d], Guilherme da Silva Lopes Fabris[d], Bruno Ipaves[d], Raphael Matozo Tromer[e], Marcelo Lopes Pereira Junior[f], Gelu Costin[g], Preeti Lata Mahapatra[a,h], Nicholas R. Glavin[i], Ajit K. Roy[i], Venkataramana Gadhamshetty[c*], Douglas Soares Galvão[d*], Alan Dalton[b*], Chandra Sekhar Tiwary[a,j*]*

[a]School of Nano Science and Technology, Indian Institute of Kharagpur, West Bengal, 721302, India.

[b]Department of Physics and Astronomy, University of Sussex, Brighton BN1 9RH, United Kingdom

[c]Department Civil and Environmental Engineering, South Dakota School of Mines and Technology, Rapid City, South Dakota 57701, United States of America.

[d]Applied Physics Department and Center for Computational Engineering & Sciences, State University of Campinas, Campinas, São Paulo 13083970, Brazil.

[e]University of Brasília, Institute of Physics, Brasilia 70910900, Federal District, Brazil.

[f]University of Brasília, College of Technology, Department of Electrical Engineering, Brasilia 70910900, Federal District, Brazil.

[g]Department of Earth, Environmental and Planetary Sciences, Rice University, Houston, TX 77005, United States of America

[h]School of Semiconductor Manufacturing, New Age Makers Institute of Technology, Gandhinagar, Gujrat, 382055, India.

[i]Materials and Manufacturing Directorate, Air Force Research Laboratory, Wright-Patterson Air Force Base, Dayton, OH, United States

[j]Metallurgical and Materials Engineering, Indian Institute of Technology Kharagpur, West Bengal, 721302, India.

**Abstract**

Two-dimensional (2D) silicates have emerged as a promising class of ultrathin materials, expanding the landscape of 2D systems beyond conventional van der Waals crystals. Their unique crystal chemistries and structural anisotropies make them attractive for applications ranging from sensors and flexoelectric devices to drug delivery and catalysis. To unlock their full potential, it is critical to understand their thickness-dependent mechanical properties within the family of 2D silicates. In this study, we investigate the nanomechanical and frictional behaviors of two structurally distinct natural silicates: layered Biotite and chain-structured Rhodonite. Using atomic force microscopy (AFM), we found that Rhodonite exhibits nearly ten times higher adhesion force and modulus response compared to Biotite. Despite this, Biotite demonstrates superior frictional performance, with ultrathin (5 nm) flakes showing a remarkably low coefficient of friction (~$0.6 \times 10^{-3}$) versus Rhodonite (~$3.6 \times 10^{-3}$). To further elucidate interlayer adhesion, density functional theory (DFT) calculations with Hubbard correction were employed. These findings offer valuable insights into the design and selection of 2D silicates for advanced mechanical and tribological applications.

**INTRODUCTION**

Atomically thin sheets have garnered tremendous interest over the last two decades, particularly within the van der Waals family. It was initially proposed that only layered materials possess the capability to be reduced to thin two-dimensional sheets, attributed to the relative ease of separation facilitated by the weak interfacial interactions between the stacked layers [1]. Subsequently, after several years, a group of researchers investigated MXenes and demonstrated that 3D nonlayered crystals could undergo chemical conversion to form layered crystals, which are amenable to exfoliation through liquid-phase exfoliation (LPE) [2,3]. The process described consists of two distinct stages: the initial phase involves the chemical etching of 13 or 14 group elements, leading to the formation of a layered structure transitioning from MAX to MXenes, followed by the exfoliation of these resultant materials [4]. Although it was not a direct exfoliation approach the original efforts in MXenes displayed thepotential for directly exfoliating non-layered crystals into nanosheets. In the following years, as researchers documented the direct exfoliation of non-layered materials, including $WO_3$ [5], $\alpha$-$Fe_2O_3$ [6], and $FeTiO_3$ [7], $MnTe$ [8], and $CaCO_3$ [9], to name a few. Silicate is emerging as a notable member within the family of complex oxides. Silicate, a material abundant in the Earth's crust, exists in both layered and non-layered arrangements. The initial preparation of the 2D silicate was conducted through synthetic methods utilizing molecular beam epitaxy (MBE) to investigate the formation and interaction of water with 2D silicate bilayers on a Pd (111) substrate [10].

Subsequently, a variety of natural silicates have been investigated for diverse applications, including pollutant removal [11], clinical uses [12], catalysis [13], enhancement of fertilization processes [14], and energy storage solutions [15], among other layered and non-layered structures for a range of applications [16]. In previous studies, Biotite, a naturally occurring layered material, was synthesized into two-dimensional sheets referred to as biotene [17]. The research focused on its potential applications in flexoelectric energy harvesting, as well as its remarkable properties as an anode for Li–Na-ion batteries [18,19]. The nonlinear absorption and optical limiting capabilities of biotene were also investigated [19]. In contrast, researchers have also explored non-layered silicates such as diopside, rhodonite, and tourmaline. Diopside has been investigated for its flexoelectric energy harvesting capabilities [20], whereas rhodonite has been examined for its applicability in gas sensing [21,22]. Tourmaline, on the other hand, has been employed as a coating on cotton substrates to support energy harvesting in wearable technologies and health monitoring systems [23]. 2D silicates are attracting considerable attention in contemporary applications owing to their inherent abundance, structural characteristics, remarkable chemical and thermal stability, as well as their varied surface chemistry [16]. The presence of hydroxyl groups and exchangeable interlayer cations in these materials facilitates relatively easy functionalization and integration into diverse matrices, rendering them highly versatile and promising for present-day applications. Therefore, it is essential to explore the intricate relationship between bonding topology and dimensionality and their impact on mechanical properties, as such insight will be beneficial for the future applicability of these materials. Furthermore, these two-dimensional materials demonstrate inherently favorable frictional properties which can facilitate the development of interfaces with exceptionally low friction [24]. This study insights into the nanomechanical and friction properties of a layered silicate, namely Biotite, in comparison to a non-layered silicate, Rhodonite. The silicates were subjected to the liquid phase exfoliation process, followed by a comprehensive characterization to determine their elemental composition, crystal structure, morphology, atomic arrangements, and the vibrational and rotational modes present. The adhesion, indentation, and modulus were measured in relation to the thickness of the nanoflakes using Peak Force-Quantitative Nanomechanical Microscopy (PF-QNM) with lateral force microscopy was utilized to evaluate the frictional characteristics of the flakes. The adhesion properties among varying layer thicknesses of biotite and rhodonite were investigated utilizing density functional theory (DFT).

## MATERIALS AND METHODS

### Synthesis

Bulk biotite (LP6) with the general formula $K(Mg,Fe)_3(AlSi_3O_{10})(OH)_2$ was obtained from a rock known as biotite pyroxenite. It was selected for establishing a geochronological standard due to its well-preserved, fresh biotite, high potassium oxide ($K_2O$) content, and an expected K–Ar age between 100 and 150 million years. This sample was sourced from the north-central region of Washington State. To isolate 2D layers from bulk biotite, a liquid phase exfoliation method was employed. A total of 10 mg of bulk biotite underwent exfoliation using a bath sonicator at ambient temperature in 50 mL of Isopropyl Alcohol (IPA). The sonication process lasted for 8 hours, after which the exfoliated sample was allowed to rest for 24 hours to monitor the precipitation at the bottom of the container. Following centrifugation, the dispersed portion of the sample was collected for subsequent analyses. Similarly, Rhodonite bulk crystals with the general formula $CaMn_3Mn(Si_5O_{15})$ sourced from Cavnic, Romania were reduced into exfoliated flakes using the same solvent. The bulk-sized crystals were first crushed into a fine powder with a mortar and pestle. This powder was subsequently mixed with 50 mL of IPA and subjected to bath sonication for 18 hours, maintaining the solvent temperature below 35 °C. After sonication, both the exfoliated sample solution was assessed for any sedimentation at the bottom of the container.

### Instrumentation

X-ray diffraction (XRD) patterns and phase identification of the sample were obtained using a Bruker D8 Advance diffractometer equipped with Cu Kα radiation (λ = 1.5406 Å), operated at 40 kV and 40 mA. Raman spectroscopic analysis was conducted at room temperature using a

WiTec UHTS 300 VIS Raman spectrometer (Germany) with a 532 nm, excitation wavelength. Surface composition and oxidation states of the exfoliated samples were analyzed using a PHI 5000 VersaProbe III scanning XPS microprobe. High-resolution transmission electron microscopy (HRTEM, Titan Themis, 300 kV) was employed to examine the topographical, compositional, and crystalline features of the material.

Quantitative nanomechanical mapping (QNM) and friction investigations were conducted using Atomic Force Microscopy (AFM). The Bruker Dimension Icon instrument was used to perform these characterizations at room temperature and 35% relative humidity. To reduce background noise and building vibrations, the AFM was placed in an insulated box above an anti-vibrant stage. Measurements were made using particular cantilevers for each mode. The test was conducted using silicon nitride probes with probe stiffness, driving frequency, and tip radii were measured at 0.4 N/m, 73.06 kHz, and 10±2 nm, respectively, to measure the change in mechanical characteristics such as modulus, adhesion, and indentation along the thickness transitioning from monolayer to bulk. Lateral force microscopy (LFM) was employed for friction mapping. For biotite, a silicon cantilever having length 210 μm, width 20 μm, and stiffness 0.15 N/m (Model: MSL-10) with a rectangular $Si_3N_4$ tip of radius 12 nm was used to quantify the lateral force by moving the tip apex in the trace and retrace direction. The silicon cantilever with a stiffness of 0.187 N/m (Model: CSG10-5) and dimensions of 241 μm for length and 33 μm for breadth was utilized for rhodonite. It should be mentioned that although though different cantilevers were used for the measurements, the force applied to the flakes was the same in both samples.

**Preparation of sample for AFM analysis**

The powdered sample of 2D biotite and rhodonite was mixed into 10 ml of IPA and ultrasonicated for up to 30 minutes. The dispersion was spin-coated over the Si wafer at 2000 rpm for 1 minute and kept for drying at 70º C for 2 hrs. The prepared sample was further used for the doing AFM analyses.

**Theory Methodology**

To simulate the exfoliation process of both, van der Waals (vdW) and non-van der Waals (non-vdW) solids, we employed Density Functional Theory (DFT) with Hubbard correction (DFT+$U$) [25], as implemented in the Spanish Initiative for Electronic Simulations with Thousands of Atoms (SIESTA) code [26,27]. Exchange-correlation effects were treated using the Perdew-Burke-Ernzerhof (PBE) functional [28] within the framework of the Generalized Gradient Approximation (GGA), combined with a double-ζ polarized (DZP) basis set composed of numerical atomic orbitals. A mesh cutoff of 150 Ry was employed for all simulations, and a Γ-centered Monkhorst-Pack k-point grid with 1×1×1 sampling was used [29]. For systems containing Fe atoms, a Hubbard $U$ parameter of 4.0 eV was applied[28]

Two distinct crystalline systems were considered: Biotite [30], a van der Waals lamellar crystal, synthesized in monolayer form and referred to as Biotene [18], and Rhodonite, a non-van der Waals material stabilized by covalent bonding [31]. The Biotite and Rhodonite unit cells contained 40 and 50 atoms, respectively.

Unit cells were incrementally stacked along the *z*-axis to construct the slab models used in the exfoliation simulations. This generated systems with $L$ = 1 to $L$ = 5 layers (e.g., mono- to penta-layer configurations) and resulted in supercells composed of approximately 200 and 250 atoms, respectively.

After stacking, a vacuum region of approximately 200 Å was inserted along the *z*-direction to suppress interactions between periodic images. These supercells comprising multiple layers and a vacuum region were fully optimized at 0 K until the residual atomic forces were below 0.05 eV/Å. Electronic convergence was considered achieved when the difference between the input and output elements of the density matrix was smaller than $10^{-4}$. Stress tensors were also evaluated at the end of each structural relaxation. Figure S8 shows the fully optimized supercells corresponding to the initial configurations for the exfoliation studies. In these images, the atomic species are represented by different colors: yellow (Si), red (O), green (Mn), purple (K), pink (Mg), and blue (Fe).

The exfoliation process was simulated by incrementally displacing the outermost layer in steps of $\delta z = 0.5$ Å, performing single-point energy calculations at each step with all atomic positions fixed. This procedure was repeated until the exfoliated layer reached a sufficient separation (~4 Å) to detach fully. After this threshold was reached, a constant separation of 30 Å was set to model the exfoliation of the subsequent slab. For each step, the total energy was recorded, and the exfoliation energy ($E_{exf}$) was computed as [32]:

$$E_{exf} = \frac{E(N-1) - E(N)}{A_{sur}}, \qquad (1)$$

where $E(N)$ and $E(N-1)$ represent the total energies of systems with $N$ and $N-1$ layers, respectively, and $A_{sur}$ is the surface area of the exfoliated monolayer. In addition to $E_{exf}$, we also evaluated the adhesion energy ($E_{ad}$), which is defined as the total energy difference between successive configurations during the layer separation process. From the resulting energy-distance curve, the adhesion force ($F_{ad}$) was estimated by computing the numerical derivative of $E_{ad}$ with respect to the displacement z:

$$F_{ad} = \frac{-dE_{ad}}{dz}. \qquad (2)$$

In the case of Rhodonite, due to Mn-O and Si-O bonds connecting adjacent layers, the exfoliation procedure inherently involves breaking covalent bonds. Therefore, the same exfoliation protocol used for Biotite was applied, but particular care was taken to monitor bond dissociation and structural deformation during each displacement step.

Before the exfoliation simulations were performed on the fully stacked supercells ($L = 5$), additional analyses were conducted to investigate the mechanical response at thicknesses different from those defined by the original unit cell (around 10 Å for Biotite and 7 Å for Rhodonite). Other structures with thicknesses of approximately 9 Å, 12 Å, and 15 Å were generated by selectively displacing partial volumes relative to the standard unit cell configuration, specifically the *c* parameter (direction of exfoliation). The exfoliation procedure was subsequently applied to these thinner slabs, enabling the evaluation of adhesion forces and structural rearrangements as a function of slab thickness. In some cases, particularly at larger thicknesses, convergence difficulties were encountered during an energy minimization, primarily due to bond-breaking events and local structural deformations induced by the exfoliation process.

**RESULTS AND DISCUSSION**

Biotite is a layered silicate that exhibits easy cleavage due to the weak van der Waals forces that bind the layers together. The structural formula for bulk biotite is $(K_{0.978}Na_{0.014})_{\Sigma=0.992}(Mg_{2.104}Fe^{2+}_{0.591}Al_{0.149}Ti_{0.079}Mn_{0.006}Fe^{3+}_{0.004}Cr_{0.003})_{\Sigma=2.916}[Si_{2.821}Al_{1.179}O_{10}](OH)_2$, comprising three bonding layers: K–O, (Si–O, Fe–O), and (MgO, Fe–O). Notably, the K–O layer serves as the separating layer during the process of cleavage. Conversely, Rhodonite is characterized as a non-layered silicate, with $SiO_4$ tetrahedra that are strongly bonded in chains along the c-axis, which are interconnected with Mn octahedra. The structural formula of rhodonite is $(Mn_{0.702} Fe^{2+}_{0.207} Ca_{0.085} Mg_{0.006} Na_{0.001})_{\Sigma=1.001} (Si_{0.999} Al_{0.001})_{\Sigma=1.00} O_3$. The detailed compositional analyses were documented in the prior research [17,21]. In order to verify the existence of these elements in the exfoliated sample, XPS measurements were performed, as illustrated in SI Figure S1 and the elemental analyses for 2D biotite and rhodonite are presented in Figures S2 and S3, respectively. The deconvoluted XPS graph indicated a significant presence of Fe, Mg, Na, K, and Ca atoms in biotite, while rhodonite exhibited a higher concentration of Mn atoms. Comparable findings were observed in the EDS mapping of the biotite and rhodonite flakes, as illustrated in SI Figure S4. **Figure 1a** illustrates the structural distinctions between biotite and rhodonite. In the case of biotite, a well-organized arrangement of layers can be observed, while rhodonite is characterized by its strongly bonded atomic structure. The morphology was further validated through scanning electron microscopy (SEM), as illustrated in **Figures 1**. **Figure 1b** clearly illustrates the layered structure of biotite, observable from both the top view and the section view, with the latter highlighted in yellow. In the case of Rhodonite, a singular crystal specimen was noted, exhibiting an absence of distinct visible layering (**Figure 1c**).

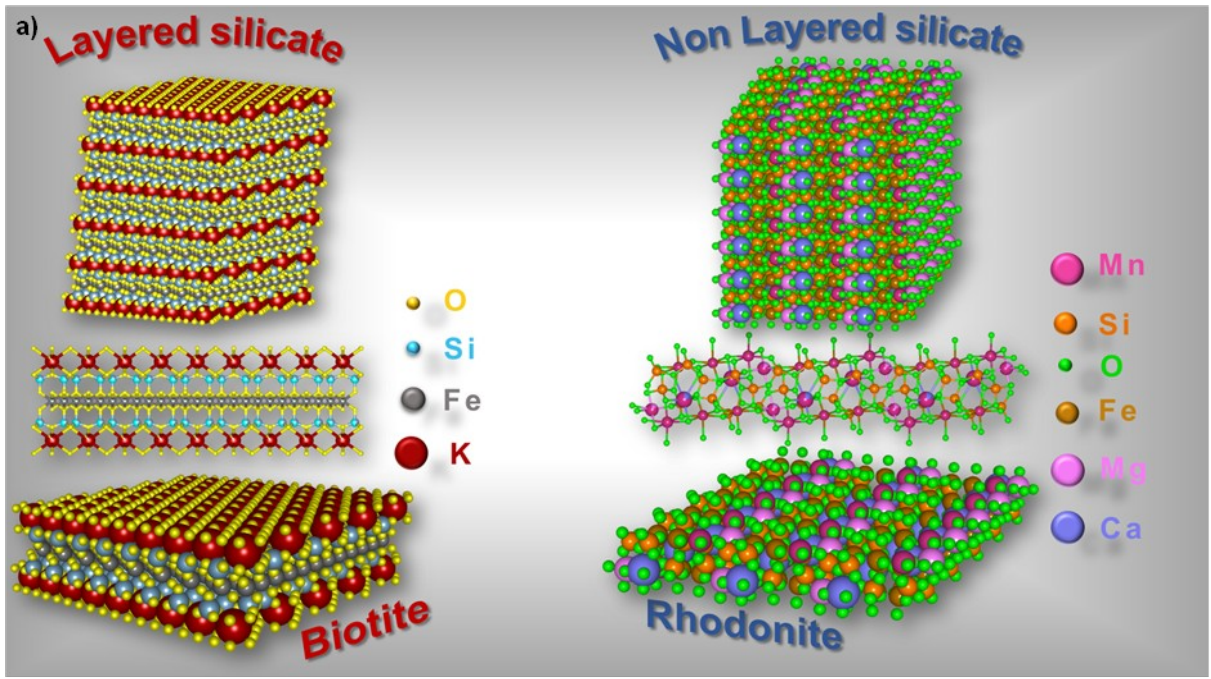

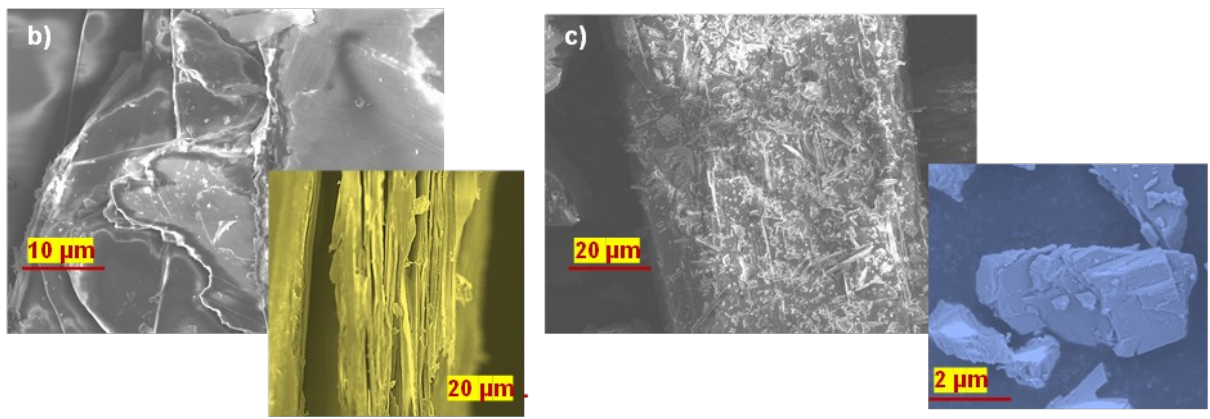

**Figure 1:** a) Schematic representation of the layered (Biotite) and non-layered (Rhodonite) silicate, b) SEM image of the bulk biotite depicting its layered structure, and c) SEM image of the bulk rhodonite.

The XRD pattern of the biotite bulk and 2D sample displayed pronounced and intense diffraction peaks, signifying a high degree of crystallinity. The significant peaks were identified at 26.51º, 35.67º, 45.03º, and 54.73º, corresponding to the (003), (131), (005), and (204) planes, respectively. In the case of its 2D counterpart, several additional peaks were noted, as illustrated in **Figure 2a**. The crystal structure exhibited a monoclinic configuration characterized by the C12/m1 space group with the cell parameters as follows: a = 5.3550 Å, b = 9.2510 Å, c = 10.2460 Å, with angles α = γ = 90° and β = 100.15° (Reference code: 96-900-1267). Conversely, rhodonite demonstrated an anorthic crystal structure characterized by the P-1 space group. These materials, due to their low symmetry and complex structure, typically display a significant number of unique crystallographic planes, resulting in multiple diffraction peaks in their XRD patterns. The most intense peaks were located at 24.86º (020), 26.5º (1$\bar{2}$0), 28.23º (201), 28.75º ($\bar{1}\bar{2}$2), 29.88º ($\bar{2}$11), 30.33º (210), 34.37º (2$\bar{2}$1), 35.61º (0$\bar{3}$2), 37.77º (220), and 41.27º (3$\bar{1}$0) as shown in **Figure 2b**. The lattice parameters for rhodonite were determined as follows: a = 6.7 Å, b = 7.68 Å, c = 11.82 Å, with angles α = 105.65°, β = 92.37°, and γ = 93.94° (Reference code: 96-900-3681). Numerous distinct peaks with the same crystal structure as the bulk were also seen in the 2D rhodonite. **Figure 2c** illustrates the Raman spectra corresponding to 2D rhodonite and biotite. The Raman spectra of 2D rhodonite exhibited a pronounced sharp peak at 670.95 cm$^{-1}$, which is associated with the bending mode of the Si-O-Si bridging bond. The observed doublet peaks at 978.6 and 1000.8 cm$^{-1}$ are ascribed to the asymmetric stretching vibrations of Si-O non-bridging bonds, which are a result of the SiO$_4$ tetrahedra present in the rhodonite structure. The cation oxygen vibration modes are observed at 421.52 cm$^{-1}$, while the lower region peaks at 331.26, 285.8, and 249.03 cm$^{-1}$ correspond to the metal-oxide bonds found in rhodonite, specifically the Mg/Fe-O bonds [33,34]. A significant intensity peak was identified at 122.39 cm$^{-1}$, attributed to the lattice vibrations of Mn-O octahedra. The intensity observed is notably untypical in silicates, potentially resulting from the presence of Mn ions on the surface of rhodonite after exfoliation. For the 2D biotite, similar peaks corresponding to the non-bridging Si-O-Si bond were detected at 947.88 and 975.76 cm$^{-1}$. The triplet peaks are associated with the vibrational modes of Si–O$_b$–Al and Si–O$_b$–Si bonds. The peaks in the lower region below 350 cm$^{-1}$ are attributed to the translational bonds of metal oxides, specifically Mg/Fe-O, while the peak near 180 cm$^{-1}$ may correspond to the K-O bond [35].

The high-resolution transmission electron microscopy (HRTEM) images of 2D biotite and rhodonite, as illustrated in **Figures 2d and 2e** respectively, provide insights into the atomic arrangement and the planes present on the surface. The insets illustrate the FFT and inverse FFT images, depicting various plane orientations towards (110) and (121) for 2D biotite, with $d_{spacing}$ of 0.37 nm and 0.33 nm, respectively, in the context of 2D biotite. In contrast, for 2D rhodonite, the plane (022) exhibits a $d_{spacing}$ of 0.33 nm.

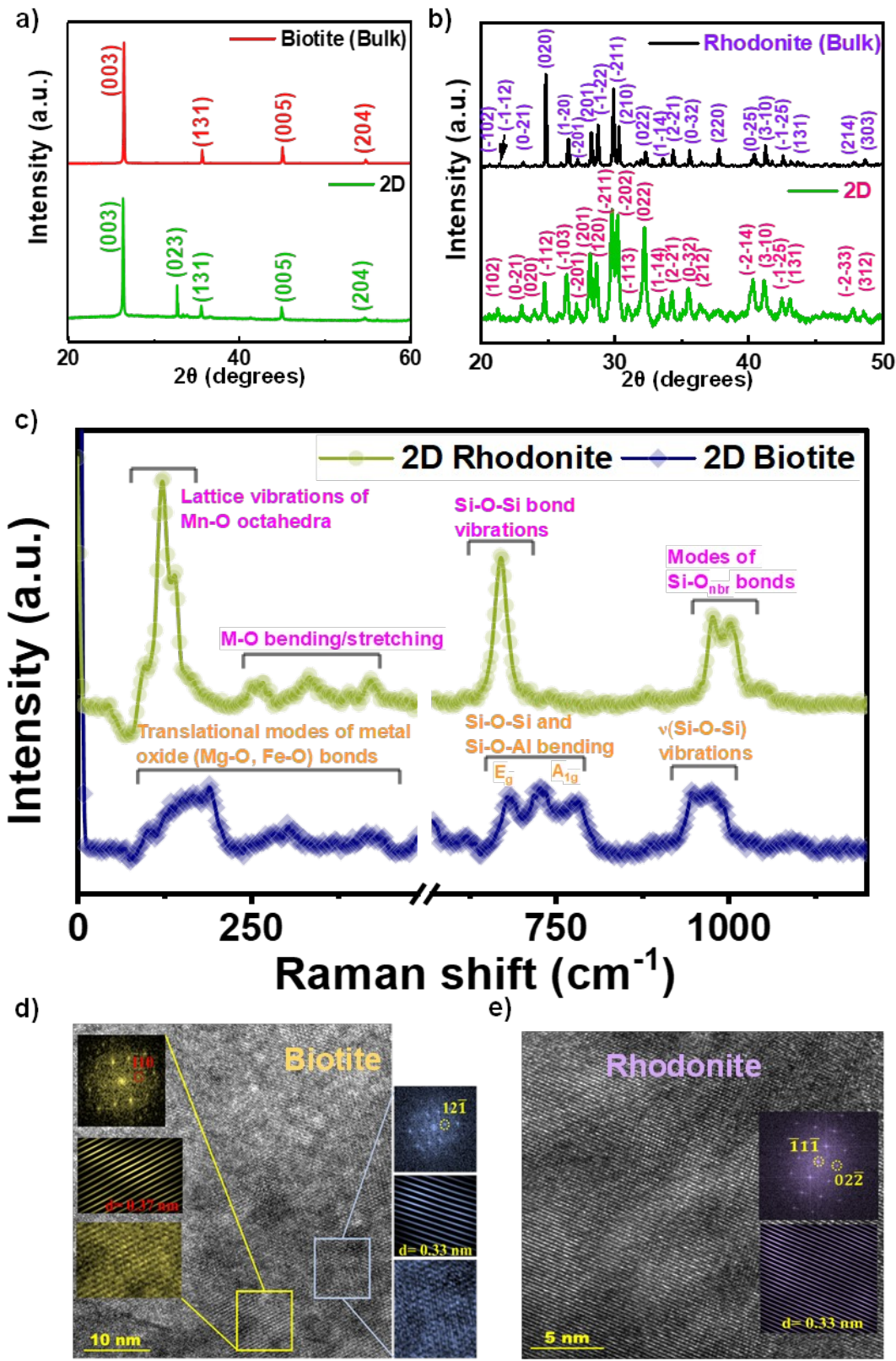

**Figure 2**: Characterizations of biotite and rhodonite. a) XRD spectra of bulk and 2D biotite, b) XRD spectra of bulk and 2D Rhodonite, c) Raman spectra of 2D biotite and rhodonite, d) TEM

analyses of 2D biotite with insets showing FFT and Inverse FFT of the different planes, and e) TEM image of 2D rhodonite with insets showing planes present on the surface.

**Mechanical Properties**

This investigation examines the nanoscale mechanical behavior of 2D Biotite and Rhodonite flakes deposited on $SiO_2$ substrates, utilizing atomic force microscopy (AFM) under similar operating conditions. As shown in **Figure 3a**, Biotite flakes exhibit an average thickness of 2 nm surrounding some flakes of higher thickness. Force-displacement spectroscopy was employed to evaluate probe–sample interactions. Retraction curves [36] corresponding to various sizes of biotite flakes as shown in SI **Figure S5a**, show a thickness-dependent decrease in adhesion force, a comparable trend also observed in Rhodonite (**SI Figure S5b**). Notably, Biotite flakes exhibit a steeper retraction slope than Rhodonite, indicating higher stiffness and better elastic recovery.

Mechanical maps and corresponding bar plots (**Figures 3b–e**) reveal that Biotite flakes have low adhesion (~2.22 ± 0.11 nN) and indentation (~2.31 ± 0.11 nm) at minimal thickness, but both increase slightly in the 5 nm range—likely due to stronger tip–sample adhesion causing deeper indentations. With further increase in thickness, both values decline, possibly due to reduced substrate influence or Biotite's compositional variability. Surface atom characteristics also play a crucial role. The calculations of the DMT modulus revealed that the low thickness value was approximately 0.75 GPa, while the values associated with increased thickness exhibited a continuous decline. The 5 nm flake, however, exhibited an atypical behavior (SI Figure S6).

Rhodonite flakes, in contrast, show markedly higher adhesion values (6.681–40 nN) as the thickness decreased (**Figures 3g and i**). Line profiles are provided in **SI Figure S7**. Unlike Biotite, Rhodonite shows increasing indentation with thickness (SI Figures S7f), which may stem from surface roughness and its non-layered crystal structure. Its DMT modulus (**Figures 3h and j**) is nearly 10 times that of Biotite, attributed to strong out-of-plane atomic bonding typical of non-van der Waals materials.

To further elucidate interlayer interactions and their influence on mechanical behavior, we conducted density functional theory (DFT) calculations on monolayer to few-layer Biotite and Rhodonite systems.

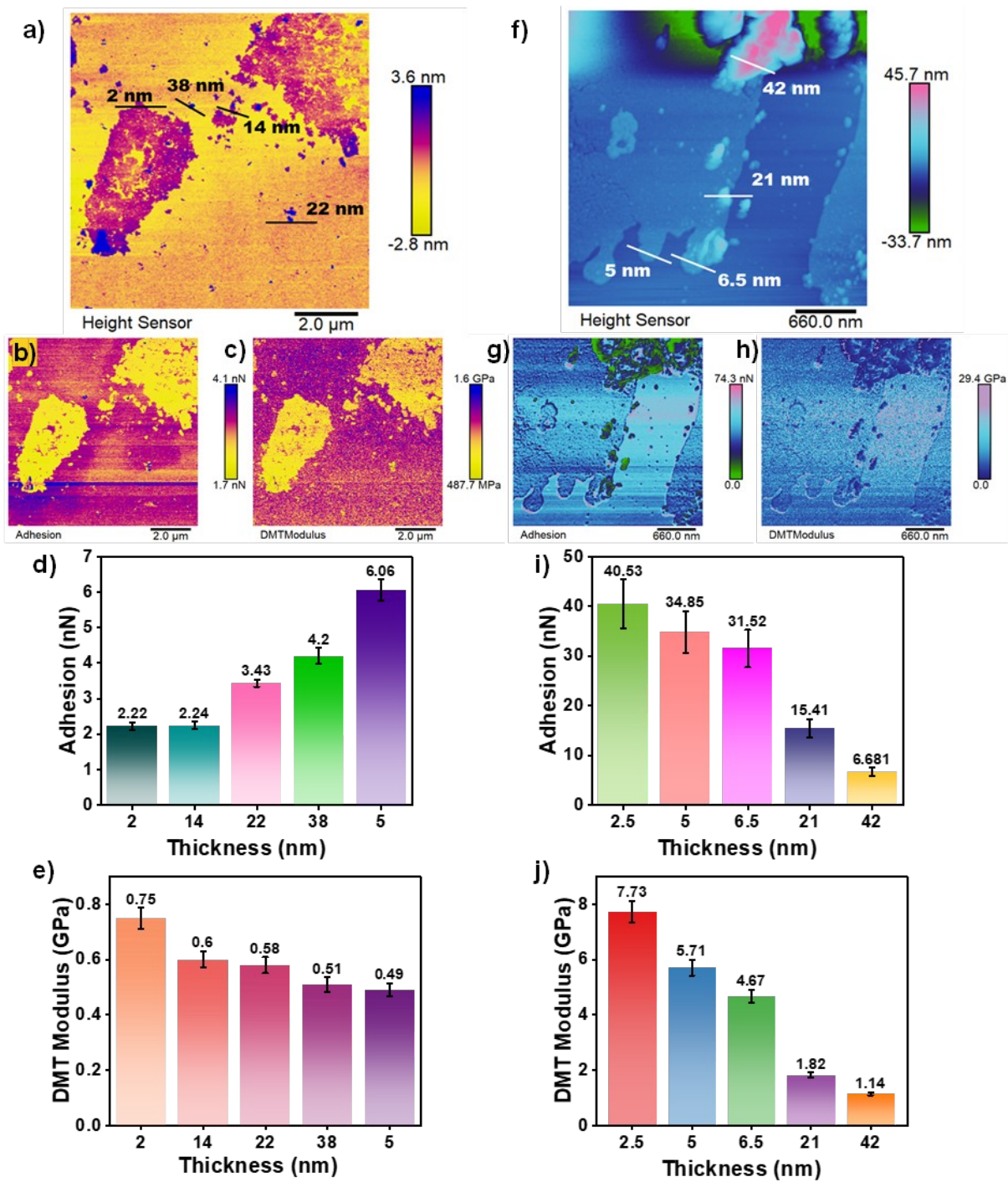

**Figure 3:** PF-QNM investigations of biotite and rhodonite flakes: a) The topographical image of the biotite flakes, b) and c) the adhesion and DMT modulus mapping of the biotite flakes, respectively, d) the bar graph depicting the relationship between adhesion and thickness of the biotite flakes, e) change in DMT modulus with increasing thickness, f) The topography image of rhodonite flakes, g) the adhesion and h) DMT modulus map of the rhodonite flakes, i) change

in adhesion force with increasing thickness rhodonite flakes and j) change in DMT modulus with increasing thickness.

**Theoretical analyses of the exfoliation behavior**

To illustrate the exfoliation procedure, **Figure 4a** presents the Biotite slab model, consisting of $L = 5$ layers with a thickness of approximately 10 Å, corresponding to the typical interlayer spacing of biotite. After complete structural optimization at 0 K, the outermost layer was selected and incrementally displaced in the out-of-plane (z-axis) direction relative to the top surface of the slab. Displacements were applied in steps of $\delta z = 0.5$ Å, up to a maximum separation of approximately 4 Å, sufficient to capture the variation in total energy during the detachment process.

**Figure 4a** shows the initial optimized configuration ($z_0$). **Figure 4b** presents the adhesion force profiles as a function of the interlayer separation for the first four exfoliated monolayers. The exfoliation process follows the methodology illustrated in SI **Figure S8 (c and d)**, involving the progressive displacement of the outermost layer along the *z*-axis. **Figure 4(b)** shows that the initial slab configuration corresponds to the system's most stable state. An increase in the system's total energy is observed upon initiating the exfoliation by displacing the top monolayer, reflecting the transition from a lower-energy to a higher-energy configuration. This trend manifests in the negative values of the adhesion force across all curves. The adhesion force gradually approaches zero for the first exfoliated monolayer as the separation increases. This indicates the effective decoupling of the monolayer from the remaining slab once a sufficiently large distance is reached. Although the force does not vanish entirely, it becomes negligible at larger separations, suggesting that the system behaves as two non-interacting subsystems. The behavior of subsequent monolayers shows the same overall tendency, as highlighted in the zoomed view in **Figure 4(c)**. While the adhesion forces for the second, third, and fourth monolayers do not reach values as close to zero as the first, their profiles remain very similar, with comparable energy minima and separation distances. This slight variation arises because the adhesion force is derived from the adsorption energy, which references different configurations depending on the exfoliated layer. For the first monolayer, the energy is evaluated concerning the original slab ($L = 5$), whereas for subsequent layers, the reference structure already incorporates previously exfoliated layers and thinner slabs. Consequently, the energy landscape subtly changes with each successive exfoliation. The overall low values of the restoring forces are a direct consequence of the lamellar nature of Biotite, where the regions connecting the layers do not exhibit significant bond formation. The absence of strong interlayer chemical bonds allows exfoliation to proceed primarily by overcoming weak vdW interactions. Nevertheless, if the exfoliation were initiated between atoms involved in stronger bonding, such as between silicon (Si) and iron (Fe) sites, bond breaking would occur, substantially increasing the energy cost of exfoliation. Such scenarios were avoided in the present simulations.

Despite the minor differences between successive monolayers, the adhesion force profiles consistently demonstrate a weak interlayer interaction, validating the adopted simulation conditions, including the vacuum region of approximately 200 Å. Although using a larger vacuum would further diminish residual interactions, the current setup accurately captures the expected behavior satisfactorily. The exfoliation energy was determined from the total energies of the initial ($z_0$) and final ($z_2$) configurations, as previously illustrated in **Figure 4a.** For the first exfoliated layer, the obtained value was 102 meV/Å$^2$. This value remained unchanged across subsequent exfoliations, further confirming the uniformity of the exfoliation process and the preservation of the structural integrity of the remaining slab. Such behavior is characteristic of materials dominated by vdW interactions, with no evidence of covalent bond breaking between adjacent layers.

We now turn to a detailed analyses of the exfoliation dynamics, focusing initially on the Rhodonite system. Visual inspection of the exfoliated structures indicates that, unlike Biotite, Rhodonite's exfoliation involves significant local structural rearrangements **(SI, Figure S9 e and f)**. This difference stems from the existence of covalent Mn-O bonds linking adjacent layers. In contrast to Biotite's vdW nature, rhodonite requires bond rupture to allow layer separation, leading to higher energetic costs and inducing localized disorder upon exfoliation. **Figure 4d** provides a closer examination of the initial rhodonite slab and the structural modifications it undergoes due to exfoliation. The exfoliation energy, computed using the same protocol as for biotite, yields an average value of 219 meV/Å$^2$ for Rhodonite, approximately 115% higher than the value found for biotite (102 meV/Å$^2$). Converting this to J/m² using the established conversion factor yields approximately 3.52 J/m². Although the absolute energy difference (ΔE) between biotite and Rhodonite is even greater when expressed in eV, the larger surface area of the rhodonite unit cell ($a$ = 7.66 Å, $b$ = 11.79 Å) compared to Biotite ($a$ = 5.34 Å, $b$ = 9.23 Å) partially mitigates this difference when normalized by area. Additionally, it is worth mentioning that Rhodonite is structurally more compact along the $c$-axis, with a lattice parameter about 40% smaller than that of biotite.

When we turn to the adhesion forces, **Figure 4e and 4f** displays the calculated force profiles for the first four exfoliated monolayers of rhodonite. The adhesion force in rhodonite is significantly larger than that observed for biotite, primarily due to the necessity of breaking strong Mn-O covalent bonds (with bond lengths around 2.10 Å) during the exfoliation process. Initially, adhesion forces as high as 17.3 nN were observed (using the conversion factor 1 eV/Å = 1.602 nN), a value approximately 10 times greater than those recorded for biotite. However, once the Mn-O bonds are broken, the adhesion force rapidly decreases toward zero, indicating that the interaction between the exfoliated monolayer and the slab ceases almost immediately.

This behavior contrasts sharply with Biotite, where long-range vdW interactions persist even at larger separations.

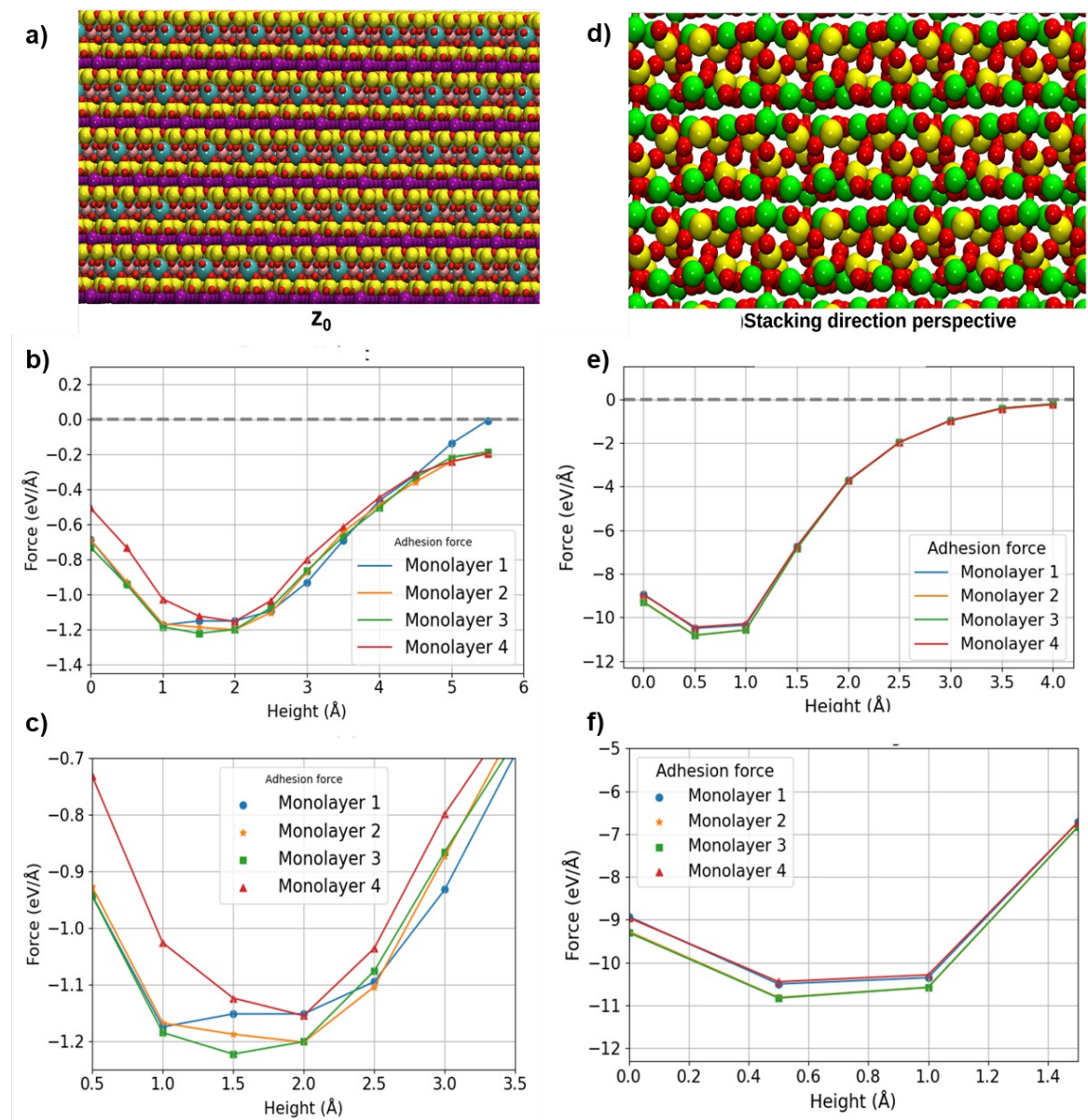

**Figure 4:** a) Illustration of the simulated exfoliation process. In (a), the Biotite slab with a thickness of $L = 5$ is shown before exfoliation. The height $z_0$ corresponds to the initial configuration, which is the optimized structure, b) Adhesion force profiles for Biotite slabs as a function of interlayer separation for the first four exfoliated monolayers highlight the progressive detachment of the layers, c) Zoomed-in view showing the similarity between the profiles of successive exfoliated layers, with comparable energy minima and recovery behaviors, d) Stacking of the pristine Rhodonite slab with $L = 5$ layers, e) Adhesion force

profiles as a function of interlayer distance for Rhodonite for the first four monolayers, and (f) zoomed view highlighting the behavior near the energy minima.

Following the previous exfoliation procedure, we analyzed different slab thicknesses for Biotite and Rhodonite to investigate the mechanical response during the separation process. Three thicknesses were considered, approximately 9 Å, 12 Å, and 15 Å, corresponding to different cuts along the out-of-plane direction, as shown in **Figure 5a**. It is essential to note that these thicknesses represent approximate values, as the actual cut must take into account the unique atomic arrangement and structural units of each crystal. In the case of Biotite, the thickness of approximately 9 Å closely matches the unit cell dimension of approximately 10 Å. In contrast, the 7 Å thickness corresponds to a cut inside the unit cell, not preserving its whole structure. Figure 4(a) presents the adhesion force profiles for Biotite slabs with different thicknesses. The force curves for thicknesses 7 Å and 9 Å show very similar behavior. This similarity is attributed to the position of potassium (K) atoms at the cleavage plane: in the 9 Å slab, the K atoms are exfoliated together with the monolayer, while in the 7 Å slab, they remain attached to the underlying structure. Since potassium atoms interact weakly with the rest of the system, their presence or absence does not significantly affect the adhesion forces. This observation is consistent with experimental results, which indicate that potassium atoms can be easily removed from biotite during the exfoliation process.

For Rhodonite, the situation is markedly different. As shown in Figure 4(b), all exfoliation processes inherently involve breaking covalent Mn-O and Si-O bonds, regardless of the selected thickness. This leads to substantially higher adhesion forces than those of biotite, highlighting Rhodonite's non-vdW nature. In this case, the cleavage corresponding to the unit cell height (~7 Å) again provides the lowest exfoliation force among the tested thicknesses, indicating a more favorable detachment pathway relative to thicker slabs.

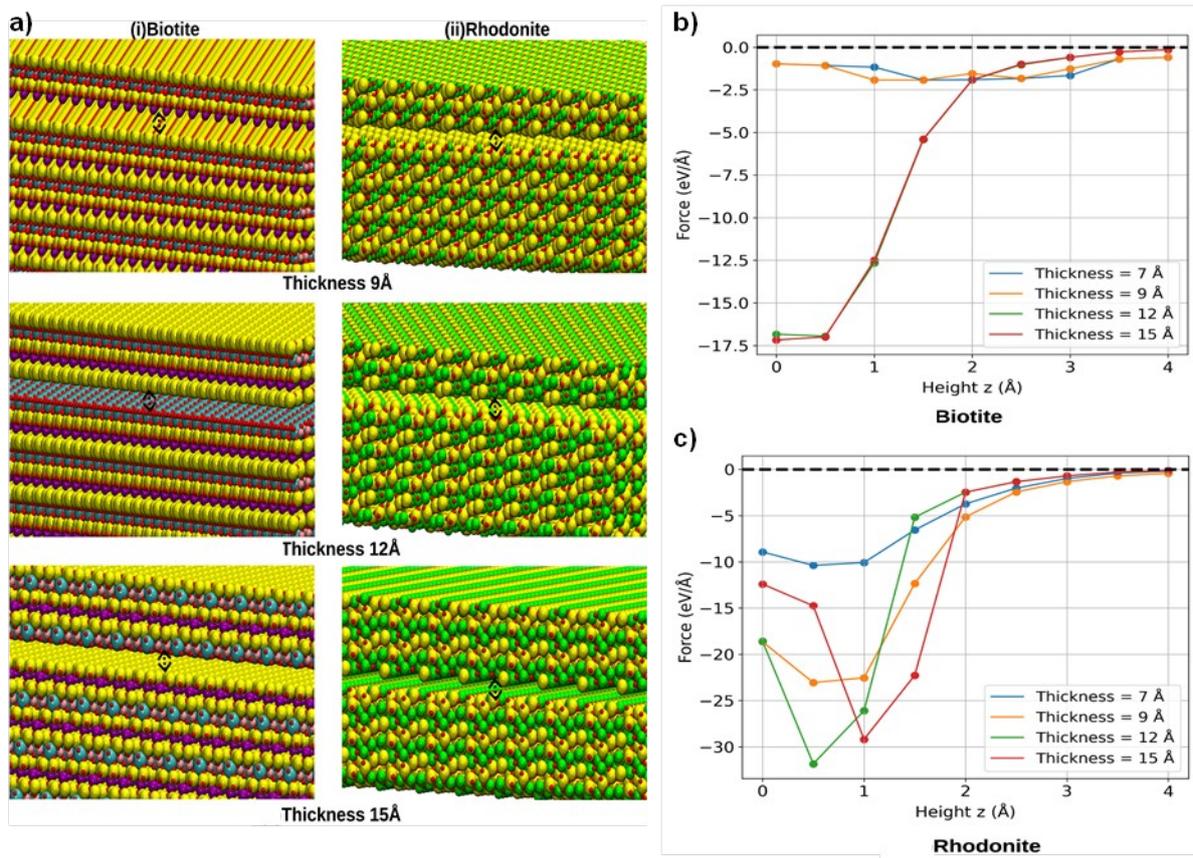

**Figure 5:** a) Atomic configurations of Biotite (i) and Rhodonite (ii) structures with three different thicknesses: 9 Å, 12 Å, and 15 Å. These models were used to investigate the adhesion response at intermediate slab thicknesses. Visual differences reveal structural rearrangements and bond modifications as the thickness increases, and Adhesion force profiles for different slab thicknesses in (b) Biotite and (c) Rhodonite. Thicknesses of approximately 7 Å, 9 Å, 12 Å, and 15 Å were considered.

Overall, the exfoliation behavior of Rhodonite differs fundamentally from that of Biotite. The need to overcome covalent bonding results in significantly higher exfoliation energies and adhesion forces, with localized structural rearrangements observable after layer separation. In the context of theoretical discussions, it has been observed that biotite, when exfoliated into individual layers, behaves as a non-interacting system. However, if exfoliation is initiated at the atomic level where stronger bonds exist, such as between silicon (Si) and iron (Fe) sites, bond breaking may take place. This may account for the varying results observed for the biotite flake with a thickness of 2.5 nm. The theoretically calculated adhesion forces were found to be greater for rhodonite in comparison to biotite flakes, thus corroborating the experimental findings. The exfoliation process in Rhodonite is characterized by considerable local structural rearrangements. The observed difference arises from the presence of covalent Mn-O bonds that connect neighbouring layers, necessitating the disruption of these bonds to achieve the formation of thin flakes.

**Frictional Studies**

The friction mapping was conducted by analysing the shear forces that occurred between the sliding tip and the surfaces of the biotite and rhodonite flakes, utilizing lateral force microscopy as the main approach of investigation [37]. The normal forces in the range of 50 nN to 300 nN were exerted on the tip to examine the alterations in friction and wear resistance of the flakes. **Figure 6a** presents the topographical representation of a biotite flake, accompanied by the friction map illustrated in **Figure 6b**. The TMR (trace minus retrace/2) values corresponding to thicknesses of 5 nm and 12 nm have been computed, as illustrated in **Figure 6c**. The study demonstrated that the friction associated with the 5 nm flake (thinner) was lower than that of the 12 nm flake (thicker), a finding that contrasts with the typical trend observed in two-dimensional materials. Thicker flakes of biotite exhibit greater stiffness out-of-plane compared to their thinner counterparts, potentially leading to a slight increase in friction response. Previous studies have observed similar behavior for graphite, when the adhesion energy between the tip and the sample exceeds the exfoliation energy, the peeling of the topmost layer results in an increase in friction due to the presence of a ripple or deformation ahead of the contact [38] The friction force shown in **Figure 6d** was determined utilizing the following formula:

$$FF = \frac{1.5 * K_t * D_s * TMR}{L * W}$$

Where FF id the Friction force, $K_t$ is the torsional spring constant, $D_s$ id the conversion factor, L is the length of the cantilever, W is the width of the cantilever and TMR is the (trace minus retrace/2) value. **Figure S10**, SI illustrates the friction maps corresponding to various applied forces. Observations indicated that a force of 123.6 nN resulted in the scratching of the flake, as illustrated by the rectangle. At an applied force of 309.06 nN, the complete delamination of the biotite flake was observed. This phenomenon may be attributed to the fact that under increased applied forces, the weak adhesion to the substrate results in the complete detachment of the flake. **Figure 6e** presents the topography of the rhodonite flakes, while **Figure 6f** illustrates the corresponding friction mapping. The data suggests that a rise in the thickness of the flakes is associated with a reduction in the TMR value, which contrasts with the behavior shown by the biotite flakes, as depicted in **Figure 6g and 6h**. The surface characteristics of the thinner rhodonite flakes may display roughness due to tearing and the presence of surface defects, alongside exfoliation processes. The applied normal forces ranged from 67.8 nN to 214 nN on the rhodonite flake, as shown in SI, **Figure S11**. The results indicated that even with an applied force of 214 nN, the flakes exhibited no signs of wear on the surface. Nonetheless, minor scratches were noted at the peripheries of the flakes. The Coefficient of friction for biotite and rhodonite flakes, each with a thickness of 5 nm, was compared as illustrated in **Figure 6i**. The coefficient of friction (CoF) for biotite flake is measured at an ultralow value of approximately 0.6 x $10^{-3}$, while rhodonite exhibits a notably higher CoF of approximately 3.6 x $10^{-3}$ at the same thickness. We compared these CoF values among various exfoliated two-dimensional materials as shown in **Figure 6j**. The CoF values exhibit a significant dependence on various factors, including the thickness of the flakes, the material of the tip, the substrate, the applied load, and ambient conditions. However, for this analyses, we utilized the previous results that were obtained with a Si/$SiO_2$ substrate for few layer thickness. The values observed for exfoliated graphene and bilayer MoS$_2$ were approximately 0.03 [39,40], indicating a close similarity between the two materials. At a load of 16 nN, Magnetene exhibited a coefficient of friction (CoF) of 0.08 [41], which is significantly higher than that of Graphene and multilayer hBN, both of which demonstrated a CoF of 0.021 [42].

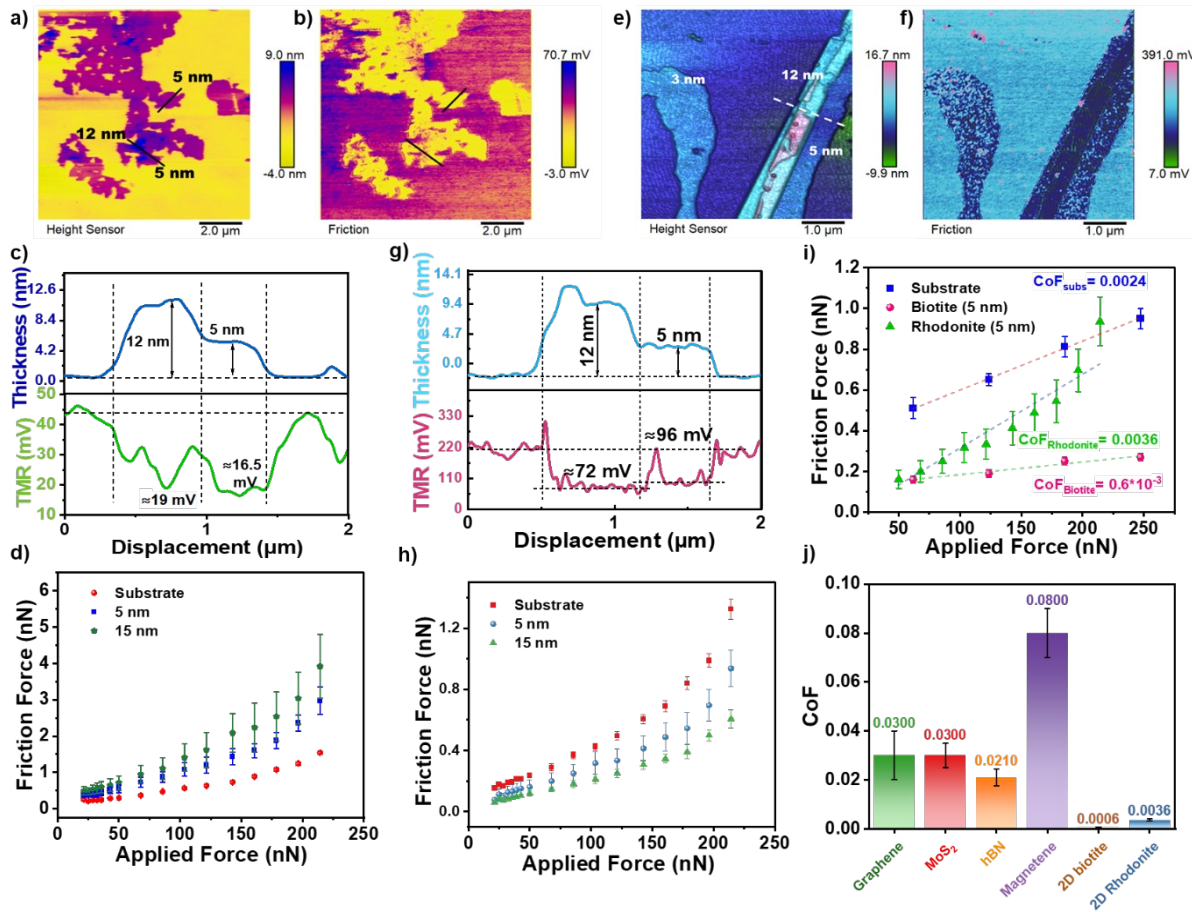

**Figure 6:** Friction studies of 2D biotite and rhodonite flakes. a) height profile of biotite flakes, b) Corresponding friction mapping of the biotite flakes, c) The height and TMR profile of 5 nm and 12 nm thick biotite flakes, d) Change in the friction force for different thickness of 2D biotite, e) Height profile of rhodonite flakes, f) the friction map of rhodonite flakes corresponding to the figure (e), g) Change in TMR values with respect to different thicknesses of rhodonite flakes, h) The resultant friction force with respect to the applied force for 5 nm and 15 nm thick rhodonite flakes, i) A comparison plot of the coefficient of friction (CoF) for the 5

nm thick biotite and rhodonite flake, and j) A comparison graph of CoF with other common 2D materials.

## CONCLUSION

This research presents a comparative analysis of two silicate minerals: layered biotite and non-layered rhodonite flakes. The DFT calculations demonstrate that biotite maintains its lamellar structural characteristics regardless of thickness. Conversely, rhodonite demonstrates gradual structural distortions and bond rearrangements, which are predominantly linked to the Mn-O and Si-O covalent bonds occurring between neighbouring layers. The observed bond ruptures may contribute to the significantly enhanced adhesion and modulus response, exhibiting a tenfold increase when compared to biotite, as demonstrated in the experimental study utilizing atomic force microscopy. A comparative analysis revealed that the biotite flake, at a thickness of 5 nm, exhibited lower frictional behavior when contrasted with the rhodonite flake. However, as the thickness increased, the friction of biotite exhibited an unusual increase, which is generally atypical for 2D materials. In contrast, the friction of rhodonite decreased with increasing thickness. The results obtained support the categorization of rhodonite as a non-van der Waals material and elucidate the difficulties encountered in its mechanical exfoliation when contrasted with layered van der Waals systems, exemplified by Biotite. These findings related to nanomechanical and tribological properties hold significant potential for applications in flexible electronics, wear-resistant coatings, and tribological interfaces in MEMS/NEMS devices, thereby contributing to long-term reliability.

**Acknowledgments**

C.S.T. acknowledges DAE Young Scientist Research Award (DAEYSRA), and AOARD (Asian Office of Aerospace Research and Development) grant no. FA2386-21-1-4014, and

Naval Research Board for funding support. C.S.T. acknowledges the funding support of AMT and Energy & Water Technologies of TMD Division of DST. The use of the EPMA facility at the Department of Earth, Environmental and Planetary Sciences, Rice University, Houston, TX, is kindly acknowledged. M.T. and A.B.D. would like to thank Sussex strategic development funds to carry out research at nanoscale.

Guilherme S. L. Fabris thanks the postdoc scholarship financed by the São Paulo Research Foundation (FAPESP) (process number 2024/03413-9) and R.B.O. thanks National Council for Scientific and Technological Development (CNPq) process numbers 151043/2024-8 and 200257/2025-0. M.L.P.J. acknowledges financial support from FAPDF (grant 00193-00001807/2023-16), CNPq (grant 444921/2024-9), and CAPES (grant 88887.005164/2024-00). Douglas S. Galvão acknowledges the Center for Computing in Engineering and Sciences at Unicamp for financial support through the FAPESP/CEPID Grant (process number 2013/08293-7). We thank the Coaraci Supercomputer for computer time (process number 2019/17874-0) and the Center for Computing in Engineering and Sciences at Unicamp (process number 2013/08293-4).

B.I. CNPq process numbers 153733/2024- 1 and FAPESP process numbers 2024/11016-0.